\documentclass[twocolumn,showpacs,preprintnumbers,amsmath,amssymb,superscriptaddress,prb,floatfix]{revtex4}

\usepackage{graphicx}
\usepackage{dcolumn}
\usepackage{bm}
\usepackage{longtable}

\begin{document}

\setlength{\parskip}{0 pt}

\preprint{APS/123-QED}

\bibliographystyle{unsrt}

\title{A high resolution, hard X-ray photoemission investigation of BaFe$_2$As$_2$: moderate influence of the surface and evidence for a low degree of Fe 3d - As 4p hybridization of the near-E$_F$ electronic states}

\author{S. de Jong}
\email{sdejong@science.uva.nl}
\affiliation{Van der Waals-Zeeman Institute, University of Amsterdam, NL-1018XE
Amsterdam, The Netherlands}
\author{Y. Huang}
\affiliation{Van der Waals-Zeeman Institute, University of Amsterdam, NL-1018XE Amsterdam, The Netherlands}
\author{R. Huisman}
\affiliation{Van der Waals-Zeeman Institute, University of Amsterdam, NL-1018XE Amsterdam, The Netherlands}
\author{F. Massee}
\affiliation{Van der Waals-Zeeman Institute, University of Amsterdam, NL-1018XE Amsterdam, The Netherlands}
\author{S. Thirupathaiah}
\affiliation{Helmholz Zentrum Berlin, Albert-Einstein-Strasse 15, 12489 Berlin, Germany}
\author{M. Gorgoi}
\affiliation{Helmholz Zentrum Berlin, Albert-Einstein-Strasse 15, 12489 Berlin, Germany}
\author{R. Follath}
\affiliation{Helmholz Zentrum Berlin, Albert-Einstein-Strasse 15, 12489 Berlin, Germany}
\author{J. B. Goedkoop}
\affiliation{Van der Waals-Zeeman Institute, University of Amsterdam, NL-1018XE
Amsterdam, The Netherlands}
\author{M. S. Golden}
\affiliation{Van der Waals-Zeeman Institute, University of Amsterdam, NL-1018XE
Amsterdam, The Netherlands}

\date{\today}

\begin{abstract}
Photoemission data taken with hard X-ray radiation on cleaved single crystals of the barium parent compound of the \textit{M}Fe$_2$As$_2$ pnictide high temperature superconductor family are presented. Making use of the increased bulk-sensitivity upon hard X-ray excitation, and comparing the results to data taken at conventional VUV photoemission excitation energies, it is shown that the BaFe$_2$As$_2$ cleavage surface provides an electrostatic environment that is slightly different to the bulk, most likely in the form of a modified Madelung potential. However, as the data argue against a different surface doping level, and the surface-related features in the spectra are by no means as dominating as seen in systems such as YBa$_2$Cu$_3$O$_x$, we can conclude that the itinerant, near-E$_F$ electronic states are almost unaffected by the existence of the cleavage surface. Furthermore, exploiting the strong changes in photoionisation cross section between the Fe and As states across the wide photon energy range employed, it is shown that the degree of energetic overlap between the iron 3d and arsenic 4p valence bands is particularly small at the Fermi level, which can only mean a very low degree of hybridization between the Fe 3d and As 4p states near and at E$_F$. Consequently, the itinerancy of the charge carriers in this group of materials involves mainly the Fe 3d - Fe 3d overlap integrals with at best a minor role for the Fe 3d - As 4p hopping parameters, and that the states which support superconductivity upon doping are essentially of Fe 3d character.

\end{abstract}


\maketitle


\section*{Introduction}
The recent discovery of an entirely new class of high $T_C$ superconductors based on (quasi) two-dimensional FeAs-layers, rather than on CuO-layers, \cite{PNoriginal} has caused significant excitement in the field of condensed matter physics. The discovery of a new family of superconductors with high transition temperatures, large critical fields and more isotropic properties than the cuprates also gives a window of opportunity in terms of future applications. Moreover, many hope that these iron pnictides can help us to gain more insight into the mechanisms that lead to unconventional superconductivity in general, or may even help unravel a now twenty year old mystery: what makes the high $T_C$ cuprates superconduct? From the beginning of the cuprate era,  surface sensitive probes such as (angle resolved) photoemission (AR)PES, and scanning tunneling spectroscopy (STS), have played an important role in determining the electronic structure of the high $T_C$ superconductors.  \cite{DamRevMod, STMoverview} Crucial pieces of the high $T_C$ puzzle have been supplied by these experimental techniques, for instance insight in the superconducting order parameter and coupling to identifiable bosonic modes. {\cite{DamRevMod} In this light, it will be no surprise that also the iron pnictides are already being studied intensively using the aforementioned techniques. \cite{LuPnic, KamPnic, DingPnic, EvPnic, BoyPnic, HoffPnic, HasPnic}

One important point to keep in mind, however, is that the surface electronic structure of a material can differ from the bulk electronic structure, in which case detailed knowledge regarding the origin and nature of these differences is required in order to fully exploit the strong points of techniques such as ARPES and STS to investigate bulk superconductivity.

The pnictides are quasi two-dimensional materials, like the cuprates, where one can assign a formal charge to each layer of the crystal structure. As the crystal symmetry at the surface is broken, one can be left with a polar surface and the possibility of having a diverging electrical field at the cleavage plane. To avoid this `polarization catastrophe', the surface of a material can be reconstructed, both structurally and/or electronically. An interesting example of the latter is thought to occur at the interface of perovskite heterostructures, such as LaAlO$_3$ / SrTiO$_3$. Such heterointerfaces are conducting, \cite{Hetero} although both oxides have a band gap of several electron volts. The diverging electrical field is quenched here by means of a partial charge transfer, giving the interface layer half the charge and opposite sign with respect to the charge of the layer below.

\begin{figure*} 
\begin{center}
\includegraphics[width=2.0\columnwidth]{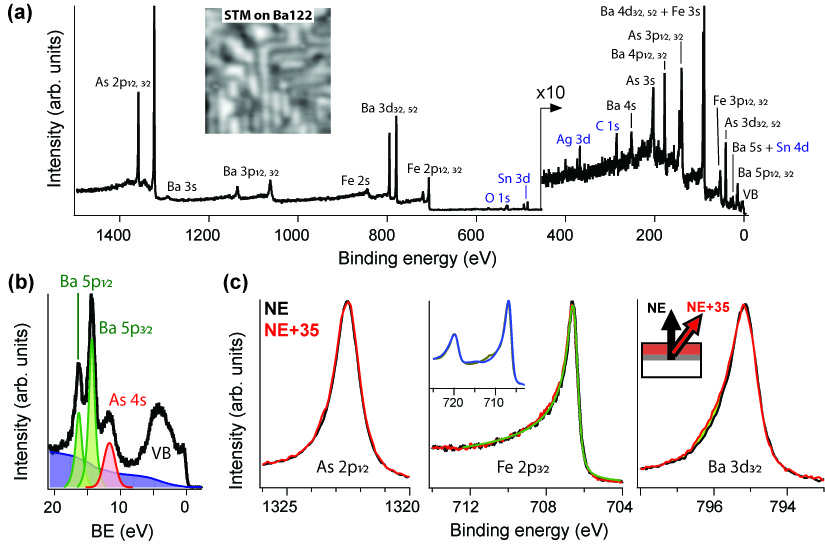}
\caption{\label{fig:pic1} Hard X-ray photoemission data from Ba122. All data taken at room temperature. (\textbf{a}) Overview spectrum taken with $h\nu=3000$~eV. The inset shows a scanning tunneling microscopy topograph ($150\times150$ \AA$^2$) from room temperature cleaved Ba122, displaying a clear surface reconstruction that lacks long range order, taken from Ref. [\onlinecite{Freek}], (\textbf{b}) Zoom of the near valence band region displaying the Ba5p (green shaded) and As4s core levels (red shaded) and an approximate inelastic background in blue. Representative core level spectra ($h\nu=2010$~eV) for all three elements are shown in (\textbf{c}). Displayed are spectra taken in a normal emission geometry (black) and 35$^{\circ}$ off normal (red), the latter decreasing the bulk sensitivity, as the mean free path length of the electron is determined by the photon energy and the escape depth thus by the emission angle (illustrated schematically in the inset of the right-most panel). The Fe 2p$_{3/2}$ peak is fitted with a single Doniac-Sunjic line shape, displayed in green. The inset to the Fe 2p$_{3/2}$ core level spectrum shows a broadened and background corrected version of this spectrum including the Fe 2p$_{1/2}$ line (blue) together with Fe2p core lines from metallic Fe(111) taken from Ref. [\onlinecite{FeFadley}] in green. Note the excellent resemblance. For the Ba 3d$_{3/2}$ line shown in the right-most panel of (c), the difference between normal and off-normal emission is highlighted in green. (color online)}
\end{center}
\end{figure*}

\begin{figure} 
\begin{center}
\includegraphics[width=1.0\columnwidth]{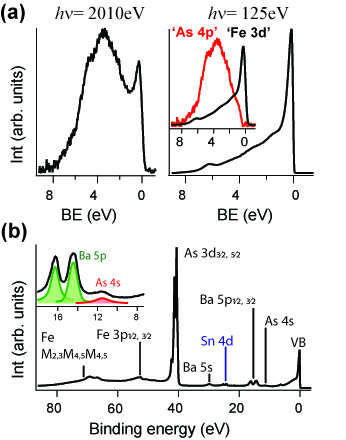}
\caption{\label{fig:pic2}Comparison of x-ray and VUV photoemission valence band data (\textbf{a}) Valence band spectra taken with $h\nu=2010$~eV, T=room temperature and $h\nu=125$~eV, T=20K [\onlinecite{footnoteT}]. The inset shows a broadened version of the $h\nu=125$~eV spectrum, representing the Fe 3d pDOS of the valence band. This curve has been subtracted from the one in the left panel to obtain the As 4p pDOS (shown as the red curve in the inset), (\textbf{b}) Shallow core levels of Ba122 recorded with $h\nu=125$~eV, T=20K. The inset shows a zoom of the Ba 5p and As core levels, with their Lorentzian peak fits (green shaded and red shaded areas, respectively). (color online)}
\end{center}
\end{figure}

In order to compare the surface and bulk properties of a material one would want to be able to tune the probing depth of the experiment. In a photoemission experiment the escape depth of the photoelectrons is sensitive to their kinetic energy, i.e. the excitation energy of the photons. Conventional (AR)PES experiments are performed with photon energies typically between 20 and 100~eV [in the (V)UV range], coinciding with a minimum in the electron escape depth and thus probing only the first few \AA ngstroms below the surface. By choosing higher excitation energies in the hard X-ray regime ($h\nu$'s of several keV), the probing depth is increased to nanometers or even tens of nanometers.

In this paper, we present results from a photoemission study of the undoped parent compound of one of the main pnictide high-T$_C$ superconductor families, BaFe$_2$As$_2$ (abbreviated forthwith as Ba122), presenting core level and $k$-integrated valence band spectra for this composition. Ba122 has been reported to become superconducting by both hole doping (substituting K for Ba) and electron doping (substituting Co for Fe), with a maximum  $T_C$ of 40~K and 22~K, respectively. \cite{KBa122, CoBa122} 

Reports in the literature have pointed out that the cleavage plane of single crystalline Ba122 is most likely to be the Ba block-layer. \cite{BoyPnic, Freek} As this compound consists of FeAs layers separated  by a \textit{single} Ba layer, this means that (in order to obtain a symmetric cleave and maintain charge neutrality) the top barium layer on the surface termination after the cleave should contain half the number of atoms compared to the bulk Ba layers. The surface of Ba122 is thus expected to differ from the bulk, structurally. This has been shown to be the case by several groups doing scanning tunneling microscopy, where a clear reconstruction of the tetragonal unit cell is visible. \cite{BoyPnic, HasPnic, Freek} The nature of the reconstruction has been reported to be 2x1 [\onlinecite{HasPnic}], or with larger period, \cite{Freek} but upon cleavage at room temperatures it lacks sufficient long range order to be observed using low energy electron diffraction. \cite{Freek} An example of such a disordered, reconstructed Ba122 surface is shown in the inset of Fig. \ref{fig:pic1}a. The question is, of course: what will be the effect of these specifically surface-related phenomena on the electronic structure of the surface and near surface region?

By comparing photoemission measurements carried out using hard X-ray radiation and data taken at conventional excitation energies, we find that the former give information mainly about the bulk electronic structure of Ba122, while the latter data show signs of an additional surface electronic structure component, representing a minor alteration of the electronic environment at the termination surface compared to the bulk. We can therefore conclude that the surface of this parent compound of the pnictide 122 high T$_C$ superconductors is electronically reconstructed with respect to the bulk electronic structure, but that the deviation from the bulk situation is modest. The implications of this altered surface structure are expected to be small for the near-E$_F$ electronic states. 

Having clarified this point, we are able to use data recorded with widely differing photon energies to show that the majority of the As 4p and Fe 3d states that make up the valence band (VB) of Ba122, possess relatively little overlap in energy. This means that the degree of hybridization between iron 3d and arsenic 4p orbitals is quite small for this compound compared to the Cu-O hybridization in the high T$_{C}$ cuprates. Importantly, the near-E$_F$ states are almost exclusively Fe 3d, thus the hopping of the itinerant charge carriers in this parent compound of the pnictide high temperature superconductors involves mainly the Fe 3d- Fe 3d overlap integrals.

\begin{table}
\caption{\label{tb:table1} Binding energies (BE) in eV of the measured core levels of Ba122, $h\nu=3000$~eV, $T=$ room temperature. The Fermi level was determined from the Fermi cut-off of piece of gold foil. The accuracy of the binding energy determination is $\pm100$~meV. The Fe 3s line coincides with the Ba 4d lines and could thus not be resolved. The Fe 3p spin-orbit splitting could also not be resolved.}
\begin{ruledtabular}
\begin{tabular} { c c c c }
\textbf{Core level} &		\textbf{BE (eV})   &		\textbf{Core level} &		\textbf{BE (eV)}\\
\hline
As 2p$_{1/2}$ &	1358.1	& 	Ba 4p$_{3/2}$				&	178.1\\
As 2p$_{3/2}$	& 1322.4	& 	As 3p$_{1/2}$				& 145.0\\
Ba 3s	& 				1292.8	& 	As 3p$_{3/2}$				& 140.1\\
Ba 3p$_{1/2}$	& 1136.0	& 	Fe 3s								& - - -\\
Ba 3p$_{3/2}$	& 1062.1	& 	Ba 4d$_{3/2}$				& 92.0\\
Fe 2s	&					 844.9	& 	Ba 4d$_{5/2}$				& 89.5\\
Ba 3d$_{3/2}$	& 795.1	& 		Fe 3p$_{1/2, 3/2}$	& 52.7\\
Ba 3d$_{5/2}$	& 779.7 &			As 3d$_{3/2}$				& 41.4\\
Fe 2p$_{1/2}$	& 719.6 &			As 3d$_{5/2}$				& 40.7\\
Fe 2p$_{3/2}$	& 706.5 & 		Ba 5s								&	29.6\\
Ba 4s	&					252.9 & 		Ba 5p$_{1/2}$				& 16.3\\
As 3s	& 				204.2 &			Ba 5p$_{3/2}$				& 14.4\\
Ba 4p$_{1/2}$ & 192.2 & 		As 4s								&	11.6\\
\end{tabular}
\end{ruledtabular}
\end{table}

\section*{Experimental}
Experiments with photon energies $h\nu=2010$ and $3000$~eV were performed at the KMC-1 beamline at the Helmholtz-Zentrum Berlin, coupled to the \textit{Scienta} R4000 analyzer of the HiKE endstation. \cite{HIKE} Experiments were carried out at room temperature in a grazing incidence geometry with a total energy resolution of 300 and 450~meV for $h\nu=2010$ and $3000$~eV, respectively, as determined from the width of the Fermi edge of a piece of gold foil. Single crystals of Ba122 larger than 1$\times$1 mm$^2$ were grown from Sn flux and cleaved in a vacuum better than $1\times10^{-9}$~mbar, resulting in shiny, flat cleavage surfaces. The level of tin impurities in the crystals was estimated from core-level intensities to be in the order of 7 atomic $\%$ and these single crystals show a magnetic and structural transition at a reduced temperature of 60~K, instead of the familiarly observed 140~K for Sn impurity-free samples. \cite{CoBa122}

Experiments with $h\nu=125$ and 140~eV were carried out at the UE112--PGMa beamline at the Helmholtz-Zentrum Berlin using an SES100 electron analyzer and were performed both at room temperature and at low temperature (25~K), with an experimental energy resolution of 30~meV. These experiments were performed on the same batch of single crystals as the HiKE experiments, and on sample surfaces cleaved both at room and at low temperature in a vacuum better than $2\times10^{-10}$~mbar. The results obtained did not depend on the temperature at which the crystal was cleaved. Spectra were taken in transmission ($k$-integrating) mode of the electron energy analyzers.

\section*{Hard X-ray data}
In Fig. \ref{fig:pic1}a an overview spectrum of Ba122 taken with $h\nu=3$~keV is shown, displaying many core level lines. The small O 1s, C 1s and Ag 3d signals come from surface contamination of the sample holder and the silver loaded epoxy that was used to attach the Ba122 crystal to the sample holder \cite{footnote-grazing}. The binding energies of identifiable core levels are listed in Table \ref{tb:table1}. Figure \ref{fig:pic1}b shows a zoom of the near-$E_F$ region with clearly distinguishable, spin orbit split Ba 5p lines at a binding energy (BE) of 15~eV, the As 4s line at 12~eV and the valence band (VB) between BE$\approx 8$~eV and $E_F$. We note that the As 4s line at 12eV has been confused in the literature with a charge transfer satellite coming from the Fe 3d states of the valence band. \cite{ding1} This spectral feature however is unlikely to originate from a satellite, as its spectral weight at $h\nu=3$~keV is comparable to the entire valence band. At lower excitation energies its relative weight becomes significantly smaller, see for instance the inset of Fig. \ref{fig:pic2}b taken with $h\nu=125$~eV. The spectral weight of the 12~eV feature traces the tabulated photoionisation As 4s cross-section values \cite{YnL} for these two photon energies, thus supporting an assignment to the As 4s shallow core level.

Figure \ref{fig:pic1}c shows spectra of representative core levels from the three elements in Ba122: As 2p, Fe 2p and Ba 3d, taken with a photon energy of 2010~eV. Comparing the line shapes of the three core levels, it is immediately clear that while the As 2p and Ba 3d peaks are quite symmetric, the Fe 2p line is not. The Fe 2p peak can be fitted with a single, Gaussian-broadened Doniac-Sunjic line shape with asymmetry parameter $\alpha=0.44$, a value that is identical to reported values for elemental Fe. \cite{FeFadley} If the asymmetry in the Ba122 Fe core lines was caused by the presence of high BE charge transfer satellites, one could expect the signature of a shoulder in the spectrum, but instead the high BE side of the core line is completely smooth. The inset in Fig. \ref{fig:pic1}c shows the measured Fe 2p spectrum together with a Fe (111) 2p line from the literature. \cite{FeFadley} The measured spectrum has been broadened with a Gaussian with full width of half maximum FWHM= 700~meV to compensate for the higher resolution of our experiment and additionally a smooth Shirley-like background has been subtracted from the hard X-ray excited data presented here so as to enable a reasonable comparison of the two spectra. One can see that the two spectra coincide perfectly. \cite{footnote2} The fact that the Ba122 Fe core level lines are so identical to those of elementary Fe means that the asymmetry of the core lines is best though of as caused by the significant partial density of states (pDOS) of Fe at the Fermi level. This enables a continuum of possible excitations during the creation of the core hole, leading to a smooth `loss-tail' at the high BE side of the core level line. In turn, the fact that the As 2p core line in Fig. \ref{fig:pic1}c is so symmetric (a fit with a Doniac-Sunjic line form yields an $\alpha$ of only 0.06), means that the As pDOS at the Fermi level is almost negligible in comparison with the Fe3d contribution. 
The core level lines were recorded both in normal emission geometry, as well as with an emission angle of $35^{\circ}$ off-normal. The latter geometry increases the surface sensitivity of the experiment, see the inset in \ref{fig:pic1}c, but yields almost identical results as normal emission for the Fe 2p and As 2p core lines. The Ba 3d line, however, is  broader when recorded in off-normal emission, as can be seen in the right-most panel of Fig. \ref{fig:pic1}c.  Although the difference is quite small, it is evident that there is a small surface contribution for the Ba core lines, causing a (modest) asymmetry in the Ba 4d line shape. 

\begin{figure} 
\begin{center}
\includegraphics[width=1.0\columnwidth]{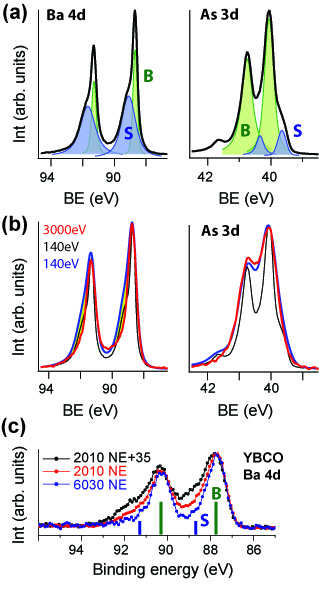}
\caption{\label{fig:pic3} Surface versus bulk electronic states of Ba122 from comparison of x-ray and VUV photoemission data (\textbf{a}) Ba 4d (left) and As 3d (right) core levels recorded with $h\nu=140$~eV (black). The surface (S), and bulk (B), contributions have been determined by a fit with 4 Gaussian broadened Lorentzians and are shown with blue and green shaded areas, respectively, (\textbf{b}) For comparison, the same core levels recorded with $h\nu=3000$~eV (red) are shown together with the data from panel (a) (thin black) and Gaussian broadened versions of the latter (blue) to compensate for the difference in experimental resolution, (\textbf{c}) Ba 4d core levels of cleaved YBa$_2$Cu$_3$O$_x$ ($x=6.75$) taken with different experimental geometries and photon energies in the hard X-ray regime. Data shown courtesy of K. Maiti and co-workers. Note the large spectral weight of the surface contribution here and the large energy difference between bulk (B) and surface (S) states, indicated in the figure with green and blue vertical lines, respectively. (color online)}
\end{center}
\end{figure}

\section*{Valence band data}
Now focusing on the near $E_F$ electronic states, a zoom of the valence band region taken with $h\nu=2010$~eV is depicted in Fig. \ref{fig:pic2}a. One can see that the total bandwidth of the valence band is about $6-8$~eV, displaying a sharp peak close to $E_F$ (the maximum being centered at about 350~meV BE) and a broad hump between 3 to 7~eV. From our core level data it is already evident that the DOS closest to $E_F$ consists primarily of Fe3d states, but at this particular photon energy ($h\nu=2010$ eV), the photoionization cross-sections (PIXs), \cite{YnL} for Fe 3d and As 4p are roughly equal, meaning that we cannot - a priori - distinguish between Fe and As pDOS. At photon energies around 100~eV however, the PIXs favor Fe 3d with respect to the As 4p states by a factor 70, making spectra measured with these photon energies representative for the Fe 3d pDOS. In Fig. \ref{fig:pic2}a one can also see a valence band spectrum recorded with $h\nu=125$~eV [\onlinecite{footnote1}], with a large peak at $E_F$ and hardly any sign of a hump at higher binding energies, showing that the Fe 3d states are indeed located close to or at $E_F$ and that the As 4p states are mainly responsible for the spectral weight between 3 and 7~eV in Fig. \ref{fig:pic2}a. The inset in Fig. \ref{fig:pic2}a shows the difference between the measured spectra in panel (a), whereby the $h\nu=125$~eV spectrum has been broadened with a Gaussian of 350~meV to account for the difference in resolution and temperature between the two measurements. This subtraction spectrum gives an estimate for the As 4p density of states. One minor remark is that the relative spectral weight of the As 4p to the Fe 3d states would then be roughly 2:1 for the hard X-ray data, judging from the area under the curves in the inset of Fig. \ref{fig:pic2}, which is a factor two too high if compared to tabulated photoionization cross-sections. These cross-sections however are listed for atomic values and it is not certain they are strictly applicable to the valence band of a material such as Ba122. The important, qualitatively robust point we make here is that there is little overlap in energy between the main maxima of the Fe 3d (350~meV) and the As 4p states (4~eV). In the literature, several studies, both theoretical \cite{Hybrid} and experimental, \cite{Lanzara} have suggested that the Fe 3d and As 4p states are strongly hybridized, thereby trying to rationalize the small measured Fe magnetic moment compared to predictions from LDA calculations. Others, however, have argued that the Fe-As overlap is small.\cite{Hybrid2} 
Although, from our data it is hard to make a quantitative statement about the hybridization between Fe and As, it is clear that electronic states between E$_F$ and the first 1.5~eV, that determine the physical properties such as magnetism and superconductivity, are (almost) exclusively of Fe 3d character. This would not mean the As 4p states are of no importance: our data would not exclude an important role for the As atoms as a source of screening of the iron on-site Coulomb repulsion. \cite{Hybrid2}  We stress here that our use of the Ba122 compound enables us to draw these conclusions regarding the Fe-As covalence in a reliable manner. Many photoemission studies in the literature have been carried out on the structurally related 1111-pnictides \cite{Lanzara, Fujimori, Koitzsch} that, besides arsenic, also contain oxygen. The O 2p states of these compounds overlap in energy with the As 4p states \cite{Hybrid2, kurmaev} and give a strong contribution to the valence band spectral weight, both at low as well as at high photon energies, thus making a trustworthy disentanglement of the As and Fe partial density of states practically impossible.

\section*{Shallow core level data}

Now zooming in on the form of the As 3d and Ba 4d levels, it is evident that for both core levels the main lines have shoulders, see panel (a) of Fig. \ref{fig:pic3}, taken with $h\nu=140$~eV. The weight of these features varies slightly from cleavage surface to cleavage surface. Note that whereas the shoulder of the As 3d line is always located at the low binding energy side of the main line, the Ba4d shoulders are reproducibly situated at the high binding energy side of the main line. The size of the binding energy shift for both elements is similar: +($-$) 400~meV for Ba (As). The appearance of these extra core level features is most naturally explained in terms of the existence of a surface and a bulk contribution. The relative weight of these contributions has been determined by fitting the spin-orbit split doublets with, in total, 4 Gaussian broadened Lorentzians, for both the As 3d and the Ba 4d core levels. We mention in passing that the spin-orbit splitting of the As 3d core level is 670~meV, which is very close to the value found for metallic arsenic, \cite{metAs} and thus we are of the opinion that this value itself cannot be taken as a signal of strong Fe - As hybridization, as was done recently in a photoemission study of a member of the 1111- pnictide family. \cite{Lanzara}

From Fig. \ref{fig:pic3}a one can see that with respect to that of As, the surface contribution for Ba is very large and significantly broadened when compared to the bulk states, which would support a picture in which the Ba atoms indeed make up the termination surface after UHV cleavage of Ba122 single crystals. The broad energy distribution of the Ba surface states might partly be due to adsorption of residual gas from the vacuum onto the surface, or, more significantly, to the absence of long range structural order in the topmost Ba layer (as shown in the inset in Fig. \ref{fig:pic1}a). \cite{HasPnic, Freek}. The fact that the surface contribution for arsenic is much more narrow than for Ba can also be rationalized within the same picture, as the `sub-surface' As layer is far less perturbed by the cleavage, and possesses a markedly reduced variation in electronic environments compared to the top-most Ba layer. Furthermore, the small spectral weight of the surface features in the As spectra compared to the Ba case can be taken as evidence that the effect of the presence of the surface is rapidly screened away for the atomic layers further below the Ba termination layer.

In Fig. \ref{fig:pic3}b the same core level spectra as in panel (a) are plotted, overlain with the same core levels measured with $h\nu=3$~keV. It is evident that the (much more surface sensitive) $h\nu=140eV$ data for Ba 4d (left panel) are much broader at the base, although they were recorded with much higher resolution than the hard X-ray data. Broadening of the VUV-excited Ba 4d data, to correct for the difference in experimental resolution, gives an asymmetric line shape with much more spectral weight at the high binding energy side than the $h\nu=3$~keV data, emphasizing the fact that the hard X-ray data is indeed probing the electronic structure representative for the bulk of Ba122. The As 3d peaks, as measured with $h\nu=140eV$, are significantly narrower compared to the X-ray data and broadening yields a spectrum that is, despite the non-negligible surface component, almost identical to the 3~keV data. This also explains why the surface contribution for As could not be resolved from comparison between the 3~keV and 2010~eV X-ray data, while it was evident for the Ba core levels. Also the Fe 3p core level (not shown) shows an altered peak form when measured with $h\nu=140eV$ compared to the hard X-ray data, although the VUV excited signal is very weak, and thus not amenable to further analysis. In addition, the small spin-orbit splitting and the Doniac-Sunjic peak form for this line disqualify a disentanglement of the surface and bulk contributions for the Fe 3p signal.

An important point to unravel is, of course, the exact origin of the difference between the surface and the bulk electronic environments of Ba122. In general, the shift of a core level $\Delta E$ can be explained by a number of terms given by the following formula:
\\
\\
$\Delta E = \Delta \mu + K \Delta Q + \Delta V_{M} -\Delta E_{R}$,
\\
\\
where $\Delta \mu$ is the change in the chemical potential, $\Delta Q$ the change in the number of valence electrons of the atom under consideration, $\Delta V_{M}$ the change in the Madelung potential, and $\Delta E_{R}$ is the change in the extra-atomic relaxation energy due to polarizability of the atoms and the conduction electrons surrounding the created core hole. \cite{hufner} In the case of Ba122 it is a non-trivial exercise to determine exactly the role played by each of these terms at the surface. One may expect, at least, that the polarization and the Madelung potential are different at the surface, although it is difficult to disentangle these two contributions. A good starting point would therefore be to compare the measured data for Ba122 to a known case from the literature, in which (electronic) surface renormalization is known to play an important role.

Let us first consider the possibility of a (grossly) different charge carrier concentration at the surface, i.e. an altered surface doping level. From photoemission investigations it is known that certain cuprates which lack a natural cleavage plane, such as YBa$_2$Cu$_3$O$_x$ ($x\approx7.0$) - YBCO for short - show such a surface doping effect. \cite{borisenkoYBCO} For YBCO this is caused by the fact that the CuO chains are ruptured upon cleavage, remaining behind - in the form of debris - on the termination surface. These chain fragments donate extra holes to the underlying copper-oxygen plane bilayer. The analogy with Ba122, where also only half of a bulk crystal layer remains on the surface after cleavage, is obvious. This type of electronic surface reconstruction for YBCO, however, has a drastic effect on the measured core level spectra, as can be seen in Fig. \ref{fig:pic3}c. Depicted are Ba 4d core levels taken, in order of decreasing bulk sensitivity, with $h\nu \approx 6$~keV and $h\nu \approx 2$~keV in normal emission geometry and $h\nu \approx 2$~keV with $35^\circ$ off-normal emission. The difference between spectra recorded with the latter two sets of experimental conditions is still very significant, showing that one has a sizable surface contribution even with an excitation energy of 2~keV. This is not the case for Ba122, Fig. \ref{fig:pic1}c, where the difference between normal and off-normal emission is nigh on indistinguishable. Also, the shift of the surface states of YBCO is quite large: about 1~eV with respect to the bulk states, more than twice that seen in Ba122. Moreover, the increased surface doping of YBCO leads to a shift of surface states toward higher binding energy both for the cations and anions (O$^{2-}$), so this mechanism can most likely be disqualified as the cause of the electronic surface reconstruction of Ba122.

The picture for Ba122 shown in Fig. \ref{fig:pic3}a is, in fact, very reminiscent of the situation seen in X-ray photoemission from the GaAs (110) surface, where the core levels are known to contain a surface contribution. \cite{GaAs1} This surface component has a shift in binding energy (with respect to the main line) that is opposite for the negatively and positively charged As and Ga ions: toward lower and higher binding energy, respectively. This shift has been explained by a change in the Madelung potential at the surface \cite{GaAs2} with a core level shift that happens to be close to what we find here for Ba122. Therefore, it is quite credible that the altered surface electronic structure of Ba122 is caused by a change in the Madelung potential, with the surface doping level itself being very close or equal to the bulk value for the Ba122 material. The question then arises as to what effect this has on the valence band states, and in particular those close to E$_F$, which are intimately involved in the superconductivity in the doped Ba122 and related materials.

While one may expect the altered surface Madelung potential to have some effect, one should bear in mind that the near-E$_F$ states in these systems are band-like and itinerant in nature, whereas the concept of a Madelung energy is more generally applicable to localized, ionic electronic levels. Secondly, in the case of GaAs (which possesses a band gap in the order of 1~eV), calculations have indicated the presence of occupied surface electronic states attributable to the altered Madelung potential in that system, lying at about 500~meV below the valence band edge. Yet, no sizable, structured contribution from these states has been identified in photoemission experiments. \cite{GaAs3} Taking all these considerations together, we conclude that the near-$E_F$ electronic structure at the surface of Ba122 is very close to that of the bulk. Furthermore, the fact that the surface doping level of Ba122 looks to be the same as the bulk (despite the presence of a reconstructed and potentially polar Ba termination layer) provides confidence that STM and ARPES should be representative probes to investigate the bulk properties of Ba122, such as spin ordering transition temperatures and superconductivity, both in terms of critical temperature and superconducting gap sizes as a function of temperature and doping.

\section*{Conclusions}
We have presented high-resolution photoemission data taken with hard X-ray and VUV photon beams on single crystals of the undoped parent compound of the electron and hole doped pnictide high temperature superconductor, BaFe$_2$As$_2$. From the line shape of the core levels it could be deduced that the near-$E_F$ electronic states are primarily of itinerant Fe 3d character. By comparing the hard X-ray excited data with results obtained using conventional VUV excitation energies, we were able to disentangle the approximate Fe and As partial densities of states in the valence band and show that the contribution of the As 4p states near the Fermi level is very small indeed. Seeing as this energy region is where the Fe 3d partial density of states is maximal, this strongly suggests that the degree of hybridization between the Fe 3d and As 4p states is minimal at and near E$_F$ for these states, a fact that could be of considerable significance in relation to the issue of the small magnetic moment found at the Fe sites in these compounds.

The exploitation of two widely differing regimes of photoelectron kinetic energies has also enabled us to examine whether the electronic structure of the cleaved Ba122 surface is the same as or close to that of the bulk. By comparison with the well studied case of GaAs, we find that the termination surface - which comprises the Ba layer of the quasi-2D crystal structure - is likely to possess a modified Madelung potential, compared to the bulk. However, the core level data do provide evidence against the existence of a surface region with differing doping level. The departure of the surface contribution to the electronic structure in Ba122 from the bulk situation is very modest in the iron pnictide. Therefore, the distorting effects of the real cleavage surface on the investigation of bulk-representative, near-$E_F$ electronic properties of Ba122 and related compounds with surface sensitive probes such as angle resolved photoemission and scanning tunneling spectroscopy are found to be minor in nature.
\\
\\
We thank J. Fink and K. Maiti for making the YBCO data available to us and for valuable discussion. We would also like to acknowledge the IFW Dresden for the provision of their SES100 spectrometer via the Helmholz Zentrum Berlin, and Huib Luigjes for expert technical support. This work is part of the research programme of the `Stichting voor Fundamenteel Onderzoek der Materie (FOM)', which is financially supported by the `Nederlandse Organisatie voor Wetenschappelijk Onderzoek (NWO)'. We also acknowledge funding from the EU (via I3 contract RII3-CT-2004-506008 at BESSY).


\begin{thebibliography}{}

\bibitem{PNoriginal} Yoichi Kamihara, Takumi Watanabe, Masahiro Hirano, and Hideo Hosono, J.Am.Chem.Soc. \textbf{130}, 3296 (2008), Ren Zhi-An, Lu Wei, Yang Jie, Yi Wei, Shen Xiao-Li, Zheng-Cai, Che Guang-Can, Dong Xiao-Li, Sun Li-Ling, Zhou Fang and Zhao Zhong-Xian, Chinese Phys. Lett. \textbf{25}, 2215 (2008).

\bibitem{DamRevMod} A. Damascelli, Z. Hussain and Z.-X. Shen, Rev. Mod. Phys. \textbf{75}, 473 (2003).

\bibitem{STMoverview} \O . Fischer, M. Kugler, I. Maggio-Aprile, C. Berthod and C. Renner, Rev. Mod. Phys. \textbf{79}, 353 (2007).

\bibitem{CuprDwave} Ding, H., A. F. Bellman, J. C. Campuzano, M. Randeria, M. R. Norman, T. Yokoya, T. Takahashi, H. Katayama-Yoshida, T. Mochiku, K. Kadowaki, G. Jennings, and G. P. Brivio, Phys. Rev. Lett. \textbf{76}, 1533 (1996).

\bibitem{LuPnic} D. H. Lu, M. Yi, S.-K. Mo, A. S. Erickson, J. Analytis, J.-H. Chu, D. J. Singh, Z. Hussain, T. H. Geballe,
I. R. Fisher and Z.-X. Shen, Nature London \textbf{455}, 81 (2008).

\bibitem{KamPnic} C. Liu, G. D. Samolyuk, Y. Lee, N. Ni, T. Kondo, A. F. Santander-Syro, S. L. Bud'ko, J. L. McChesney, E. Rotenberg, T. Valla, A. V. Fedorov, P. C. Canfield, B. N. Harmon and A. Kaminski, Phys. Rev. Lett. \textbf{101}, 177005 (2008).

\bibitem{DingPnic} H. Ding, P. Richard, K. Nakayama, K. Sugawara, T. Arakane, Y. Sekiba, A. Takayama, S. Souma, T. Sato, T. Takahashi, Z. Wang, X. Dai, Z. Fang, G. F. Chen, J. L. Luo, and N. L. Wang, Europ. Phys. Lett \textbf{83}, (2008).

\bibitem{EvPnic} D. V. Evtushinsky, D. S. Inosov, V. B. Zabolotnyy, A.Koitzsch, M. Knupfer, B. Buchner, G. L. Sun, V. Hinkov, A. V. Boris, C. T. Lin, B. Keimer, A. Varykhalov, A. A. Kordyuk and S. V. Borisenko, arXiv:0809.4455v1 [cond-mat.supr-con] (2008)

\bibitem{BoyPnic} M. C. Boyer, Kamalesh Chatterjee, W. D. Wise, G. F. Chen, J. L. Luo, N. L. Wang and E. W. Hudson, arXiv:0806.4400v2 [cond-mat.supr-con] (2008).

\bibitem{HoffPnic} Yi Yin, M. Zech, T. L. Williams, X. F. Wang, G. Wu, X. H. Chen, J. E. Hoffman, arXiv:0810.1048v1 [cond-mat.supr-con] (2008).

\bibitem{HasPnic} D. Hsieh, Y. Xia, L. Wray, D. Qian, K. Gomes, A. Yazdani, G.F. Chen, J.L. Luo, N.L. Wang and M.Z. Hasan, arXiv:0812.2289v1 [cond-mat.supr-con] (2008).

\bibitem{Hetero} A. Ohtomo and H.Y. Hwang, Nature London \textbf{427}, 423 (2004).

\bibitem{KBa122} H. Chen, Y. Ren, Y. Qiu, Wei Bao, R. H. Liu, G. Wu, T. Wu, Y. L. Xie, X. F. Wang, Q. Huang and X. H. Chen, arXiv:0807.3950v1 [cond-mat.supr-con] (2008).

\bibitem{CoBa122} Jiun-Haw Chu, James G. Analytis, Chris Kucharczyk and Ian R. Fisher, arXiv:0811.2463v1 [cond-mat.supr-con] (2008).

\bibitem{Freek} F. Massee, Y. Huang, R. Huisman, S. de Jong, M. S. Golden, and J. B. Goedkoop, arXiv:0812.4536v1 [cond-mat.supr-con] (2008).

\bibitem{HIKE} F. Scheafers, M. Mertin and M. Gorgoi, Rev. Sci. Instrum. \textbf{78}, 123102 (2007).

\bibitem{footnote-grazing} Due to the extreme grazing incidence geometry necessary to better match the inelastic mean-free path length of the photoelectrons and the penetration depth of the hard X-rays, it is unavoidable to also pick up signals from the sample mounting.

\bibitem{ding1} H. Ding, K. Nakayama, P. Richard, S. Souma, T. Sato, T. Takahashi, M. Neupane, Y.-M. Xu, Z.-H. Pan, A.V. Federov, Z. Wang, X. Dai, Z. Fang, G.F. Chen, J.L. Luo and N.L. Wang, arXiv:0812.0534v1 [cond-mat.supr-con] (2008).

\bibitem{YnL} J. J. Yeh and I. Lindau, Atomic Data and Nuclear Data Tables \textbf{32}, l (1985). These listed values, used throughout this document, are values for the \textit{atomic} PIXs and are not necessarily directly applicable to more complicated compounds, but can be well used as an estimate to the real values.

\bibitem{FeFadley} A. Fanelsa, R. Schellenberg, F. U. Hillebrecht, E. Kisker, A. P. Kaduwela, C. S. Fadley and M. A. Van Hove, Phys. Rev. B \textbf{54}, 17962 (1996).

\bibitem{footnote2} The literature spectrum of Fe (111) shown in Fig. \ref{fig:pic1}c has a small satellite around a binding energy of 712~eV, which is due to the fact that this data was taken with a Mg\textit{k$\alpha$} lab source. This feature is thus not a property of the Fe (111) Fe 2p core level line shape.

\bibitem{footnoteT} Although the hard X-ray and VUV valence band spectra in Fig. \ref{fig:pic2}a are taken with different temperatures, it was checked and confirmed that data taken at room temperature with VUV radiation did not differ significantly from the low temperature data.

\bibitem{footnote1} We note that, due to the small angular acceptance of the SES100 analyzer, the $k$-space integration of the $h\nu=125$~eV spectra is over only part of the 2D Brillouin zone. The depicted spectra are integrated over $40\%$ of the Brilllouin zone area around $\Gamma$. Using a polar rotation of the sample it was checked and confirmed that integration around different parts of the Brillouin zone, for instance $(\pi,0)$, yielded very similar spectra.

\bibitem{Hybrid} J. Wu, Ph. Phillips and A. H. Castro Neto, Phys. Rev. Lett. \textbf{101}, 126401 (2008).

\bibitem{Lanzara} D.R. Garcia, C. Jozwiak, C.G. Hwang, A. Fedorov, S.M. Hanrahan, S.D. Wilson,
C.R. Rotundu, B.K. Freelon, R.J. Birgeneau, E. Bourret-Courchesne, and A. Lanzara, arXiv:0810.3034v1 [cond-mat.supr-con] (2008).

\bibitem{Hybrid2} G.A. Sawatzky, I.S. Elfimov, J. van den Brink and J. Zaanen, arXiv:0808.1390v1 [cond-mat.supr-con] (2008).

\bibitem{Fujimori} W. Malaeb, T. Yoshida, T. Kataoka, A. Fujimori, M. Kubota, K. Ono, H. Usui, K. Kuroki, R. Arita, H. Aoki, Y. Kamihara, M. Hirano, H. Hosono, arXiv:0806.3860v1 [cond-mat.supr-con] (2008).

\bibitem{Koitzsch} A. Koitzsch, D. Inosov, J. Fink, M. Knupfer, H.Eschrig, S. V. Borisenko, G. Behr, A. K\"ohler, J. Werner, B. B\"uchner, R. Follath and H. A. D\"urr, Phys. Rev. B \textbf{78} ,180506(R).


\bibitem{kurmaev} E. Z. Kurmaev, R. G. Wilks, A. Moewes, N. A. Skorikov, Yu. A. Izyumov, L. D. Finkelstein, R. H. Li and X. H. Chen, arXiv:0805.0668v3 [cond-mat.supr-con] (2008).

\bibitem{metAs} M. K. Bahl and R. L. Watson, Surface Science \textbf{54}, 540 (1976).

\bibitem{hufner} S. H\"ufner, \textit{Photoelectron Spectroscopy} (Springer-Verlag, Berlin), 1995.

\bibitem{borisenkoYBCO} M. C. Schabel, C.-H. Park, A. Matsuura, Z.-X. Shen, D. A. Bonn, Ruixing Liang, and W. N. Hardy, Phys. Rev. B \textbf{57}, 6090 (1998), V. B. Zabolotnyy, S. V. Borisenko, A. A. Kordyuk, J. Geck, D. S. Inosov, A. Koitzsch, J. Fink, M. Knupfer,
B. Büchner, S.-L. Drechsler, H. Berger, A. Erb, M. Lambacher, L. Patthey, V. Hinkov and B. Keimer, Phys. Rev. B \textbf{76}, 064519 (2007), M. A. Hossian, J. D. F. Mottershead, D. Fournier, A. Botswick, J. L. McChesney,
E. Rotenberg, R. Liang, W. N. Hardy, G. A. Sawatzky, I. S. Elfimov, D. A. Bonn and A. Damascelli, Nature Phys. \textbf{4}, 527 (2008).

\bibitem{GaAs1} D. E. Eastman, T.-C. Chiang, P. Heimann and F. J. Himpsel, Phys. Rev. Lett. \textbf{45}, 656 (1980).

\bibitem{GaAs2}A. B. McLean, Surface Science Lett. \textbf{220}, 671 (1989), J.W. Davenport, R.E. Watson, M.L. Perlman and T.K. Sham, Solid State Commun. \textbf{40}, 999 (1981), W. M\"onch, Solid State Commun. \textbf{58}, 215 (1986).

\bibitem{GaAs3} P. E. Gregory and W. E. Spicer, Phys. Rev. B \textbf{13}, 725 (1976), D. J. Chadi, Phys. Rev. B \textbf{18}, 1800 (1978).

\end{thebibliography}
\end{document}